\documentclass{article}
\usepackage{amsmath}
\usepackage{amsfonts}
\usepackage[symbol]{footmisc}
\usepackage[]{hyperref}
\newcommand{\onlinecite}[1]{\cite{#1}}
\newcommand{\mycite}[1]{~\cite{#1}}
\newcommand{\runninghead}[2]{}
\newcommand{\address}[1]{}
\newcommand{\D}{\,\mathrm{d}}
\newcommand{\pd}[2]{\ensuremath{\frac{\partial #1}{\partial #2}}}
\newcommand{\pdl}[2]{\ensuremath{\partial #1 / \partial #2}}
\newcommand{\ie}{\textit{i.e.},~}

\newcommand{\vs}{\ensuremath{v_\text{s}}}
\newcommand{\tvs}{\ensuremath{\tilde{v}_\text{s}}}
\newcommand{\rs}{\ensuremath{\rho_\text{s}}}
\newcommand{\rc}{\ensuremath{r_\text{c}}}
\newcommand{\rn}{\ensuremath{\rho_\text{n}}}
\newcommand{\vn}{\ensuremath{v_\text{n}}}

\newcommand{\z}{\zeta}
\DeclareMathOperator{\Imp}{Im}
\renewcommand{\tau}{t}
\author{L.A.\,Melnikovsky\footnote{E-mail: leva@kapitza.ras.ru}}
\address{
P.L.\,Kapitza Institute for Physical Problems\\
Russian Academy of Sciences, 119334 Moscow, Russia}
\title{Heat Transfer in Superfluids: Effect of Gravity}
\date{}
\runninghead{L.A.\,Melnikovsky}{Heat Transfer in Superfluids}
\begin{document}
\maketitle
\begin{abstract}
  We discuss the influence of an external field on energy transport in
  superfluid. He-II is not isothermal in presence of Earth gravity;
  instead, it supports finite temperature gradient given by a
  Fourier-like equation. We calculate asymptotic behavior of the
  effective heat resistance in the vicinity of the
  \mbox{$\lambda$-transition}.
\end{abstract}

\section{INTRODUCTION}
Superfluid is known to be (and it was originally referred to as) a
\textit{super-heat-conductor}. The heat in superfluid is transfered by
the normal flow. This is fundamentally different\mycite{LL6} from the
Fourier law thermal conduction in ordinary liquids. Instead, it has a
nature of \textit{convection}. In uniform superfluid the flow (and the
heat transfer) is dissipationless, \ie it happens at constant
temperature.

Experimentally observed dissipation in bulk uniform superfluid (\ie
when the friction with walls is neglected) is believed to be due to
the vortex creation. This scenario leads to the temperature gradients
with complicated nonlinear dependence on the supported heat flux. We
show that gravity can also explain this dissipation.

Earth gravity, like any external field, destroys the system
homogeneity: it produces a pressure gradient across the sample. This
inhomogeneity is a cause of the normal flow dissipation. In other
words, gravity is a friction mechanism between the normal flow and Earth.
Due to the friction, the liquid holds a finite temperature gradient
whose leading term is proportional to the energy flux. Unlike natural
convection in ordinary fluids, this scenario of energy transfer is
described by a Fourier-like equation.

To investigate the effect quantitatively one needs to correct
superfluid hydrodynamic equations by adding external field.  In
Sec.\ref{hydro-sec} we write down these equations. The form of
dissipative terms however is not self-evident. To ascertain
correct appearance of the equations we use Landau-Khalatnikov
superfluid hydrodynamic theory in a free-fall frame of reference.
Translating all terms to the laboratory frame we unambiguously get
sought equations.

In Sec.\ref{effe-heat} we apply obtained equations to the problem of
energy transfer in superfluid. For simplicity we restrict
ourselves to one-dimensional vertical setup where influence of walls
is neglected and the energy goes either up or down. With the help of
linearized hydrodynamic equations we find temperature gradient induced
by the heat flux. In linear regime the flux and the gradient are
proportional to each other. We conclude the section by an expression
for the effective heat resistance of this system.

Effective heat resistance derived in Sec.\ref{effe-heat} becomes
larger in the vicinity of the critical temperature. In
Sec.\ref{crit-sec} we show that it diverges at $\lambda$-point and
calculate its critical behavior.

\section{HYDRODYNAMICS}
\label{hydro-sec}
Ordinary superfluid hydrodynamic equations are modified by external
fields. To find out the effect of gravity we use the equivalence
principle of general relativity. For the uniform gravity field of
Earth we define a non-inertial free-fall frame of reference
$\tilde{K}$. It has constant acceleration $\mathbf{g}$ with respect to
the laboratory frame $K(t,x^i)$. We also demand that at $t=0$ this frame has
zero velocity relative to $K$. This ensures that hydrodynamic
variables are equal in two frames.  In the frame $\tilde{K}$ no
gravity is present and (in Newtonian limit) the superfluid is governed by
conventional equations\mycite{khalat}
\begin{equation}
\label{superhydro}
\left\{
\begin{aligned}
  \pd{\tilde{\rho}}{t}  &=- \pd{\tilde{j}^i}{x^i}      \\
  \pd{\tilde{j}^k}{t}   &=- \pd{\tilde{\Pi}^{ik}}{x^i} \\
  \pd{\tvs^k}{t}        &=- \pd{\tilde{\Phi}}{x^k}     \\
  \pd{\tilde{E}}{t}     &=- \pd{\tilde{Q}^i}{x^i}     ,
\end{aligned}
\right.
\end{equation}
where tilde attributes quantities to $\tilde{K}$, $\rho$ is the fluid
density, $\mathbf{j}$ is the mass flux, $\mathbf{\vs}$ is the
superfluid flow velocity, and $E$ is the energy density. The fluxes in
\eqref{superhydro} are given by
\begin{equation}
  \tilde{\Pi}^{ik} =
                     \vs^i j^k
                     +\vn^k j^i
                     -\rho \vn^k \vs^i
                    +p\delta_{ik}
                    +\tilde{\pi}^{ik},
\end{equation}
\begin{equation}
\tilde{\pi}^{ik}=
                    -\eta
                     \left(
                       \pd{\vn^i}{x^k} + \pd{\vn^k}{x^i} 
                       - \frac{2}{3}\pd{\vn^l}{x^l}\delta^{ik}
                     \right)
                     -\delta^{ik}
                     \left(
                       \z_1 \pd{(j^i-\rho\vn^i)}{x^i}+
                       \z_2 \pd{\vn^l}{x^l}
                     \right),
\end{equation}
\begin{equation}
  \tilde{\Phi} =
                     \mu + \vs^2/2 + \tilde{\phi},
\end{equation}
\begin{equation}
\tilde{\phi}=
                     -\z_3\pd{(j^i-\rho\vn^i)}{x^i}
                     -\z_1\pd{\vn^i}{x^i},
\end{equation}
and
\begin{equation}
  \tilde{Q}^{i} =
                     \left(\mu+\frac{\vs^2}{2}\right)j^i
                    +ST\vn^i
                    +\vn^i \vn^k \left(j^k -\rho\vs^k  \right)
                    +(j^i-\rho\vn^i)\tilde{\phi}
                    +\vn^k \tilde{\pi}^{ik} 
                    -\kappa \pd{T}{x^i}.
\end{equation}
Here $\mathbf{\vn}$ is the velocity of the normal flow, $p$ is the pressure,
$\mu$ is the chemical potential, $S$ is the entropy density, $T$ is
the temperature, and $\eta$, $\z_1$, $\z_2$, $\z_3$, $\kappa$ are kinetic
coefficients.  As usual, we took not all the dissipative terms that
are in principle possible, but only the largest. In general
the dissipation in isotropic superfluid
is described by 13 kinetic coefficients 
rather than by 5 (A.~Clark 1963, see discussion in Ref.\onlinecite{LL6}, p.525).

To rewrite the equations in laboratory frame of reference one must
translate left-hand side terms in \eqref{superhydro} to $K$. We
finally get
\begin{equation}
\label{Gsuperhydro}
\left\{
\begin{aligned}
  \pd{\rho}{t}  + \pd{j^i}{x^i}      & = 0 \\
  \pd{j^k}{t}   + \pd{\tilde{\Pi}^{ik}}{x^i} & = \rho g^k \\
  \pd{\vs^k}{t} + \pd{\tilde{\Phi}}{x^k}     & = g^k \\
  \pd{E}{t}     + \pd{\tilde{Q}^i}{x^i}      & = j^k g^k.
\end{aligned}
\right.
\end{equation}

\section{EFFECTIVE HEAT RESISTANCE}
\label{effe-heat}
For the steady-state flow all time derivatives in \eqref{Gsuperhydro}
vanish. Consider one-dimensional system where all
velocities and fluxes are vertical. Let $z$-axis run down along
$\mathbf{g}$. In typical heat transfer experiment\mycite{exp1} the energy
is carried by the superfluid counterflow, where normal and superfluid
velocities are directed oppositely to each other to keep zero net mass
flux. The condition $j=0$ is satisfied along the height of the cell.
Under these circumstances the first equation in \eqref{Gsuperhydro} is
trivially satisfied and the other three have the form
\begin{equation}
\label{time-derivatives}
\left\{
\begin{aligned}
\Pi'  &= \rho g \\
\phi' &= g \\
Q'    &= 0,
\end{aligned}
\right.
\end{equation}
where dashes denote the \pdl{}{z} derivative.

Within linear (on velocity and temperature gradient) approximation the
fluxes in \eqref{time-derivatives} are given by
\begin{equation}
\label{flux-pi}
  \Pi  = p  + \z_1 (\rho\vn)'-\z_2\vn'-\frac{4}{3}\eta \vn',
\end{equation}
\begin{equation}
\label{flux-phi}
  \phi = \mu+\z_3(\rho\vn)' - \z_1 \vn',
\end{equation}
\begin{equation}
\label{flux-q}
    Q  = TS\vn - \kappa T'.
\end{equation}

Keeping lowest nontrivial gradient terms (thus assuming the effect of
$\mathbf{g}$ to be small) and excluding $\mu'$ by means of thermodynamic
identity \mbox{$\rho\D\mu=\D p - S\D T
$}, we obtain from \eqref{time-derivatives}, \eqref{flux-pi} and \eqref{flux-phi}
\begin{equation*}
A'\vn'+
B  \vn +
ST'=0,
\end{equation*}
where we introduced
\begin{equation*}
  A=2\rho\z_1 -\frac{4}{3}\eta - \z_2 - \rho^2\z_3,
\end{equation*}
\begin{equation*}
  B=
  \left(\z_1'-\rho\z_3'\right)\rho' +
  \left(\z_1 -\rho\z_3 \right)\rho''.
\end{equation*}
Finally employing \eqref{time-derivatives} and \eqref{flux-q} we obtain a Fourier-like equation
\begin{equation}
\label{fourier}
  T'=-\varrho Q,
\end{equation}
where the effective heat resistance $\varrho$ is determined by
\begin{equation}
\label{rho}
\varrho=
\frac{1}{TS^2}\left(B-\frac{S'}{S}A' 
\right).
\end{equation}

We see that in linear approximation the temperature gradient is
proportional to the heat flux and does not depend on its direction.
The effective heat resistance is of the second order in
system inhomogeneity, \ie $\varrho\propto g^2$. If higher terms are
taken into account $\varrho$ will be different when heated from above
or from below.

\section{CRITICAL BEHAVIOR}
\label{crit-sec}
The effect discussed in previous section will be present in any
Earth-based energy transport experiment, but it seems to be beyond
experimental accuracy unless measured in the vicinity of the
$\lambda$-point. To find critical behavior of the effective heat
resistance we need to estimate kinetic coefficients entering
\eqref{rho}.

This can be done using the ideas of dynamic scaling\mycite{PP}.  The
specific heat index for helium is very small\mycite{alpha}, we further
let it equal to zero. The superfluid density \rs\ follows Josephson
scaling relation\mycite{joseph} $\rs \propto \rc^{-1} \propto
\tau^{2/3}$, where \rc\ is the correlation
length and $\tau$ is the reduced temperature
$\tau=T_\lambda-T$.\footnote{The same letter $t$ is used to represent
the time and the reduced temperature. 
This should not lead to confusion however: the meaning is
clear from the context.}

Second sound damping\mycite{khalat} at the boundary of the critical
region is inversely proportional to the correlation length
\begin{equation}
\label{damping}
\rc^{-1} \propto
  \Imp \frac{\omega}{u_2} = \frac{\omega^2}{2\rho u_2^3}
  \left(
    \frac{\rs}{\rn}
    \left(\frac{4}{3}\eta+\z_2+\rho^2\z_3-2\rho\z_1\right)
    +\frac{\kappa\rho}{T}\frac{\partial T}{\partial S}
  \right).
\end{equation}
Here $\rn=\rho-\rs$, $\omega$ and $u_2$ are the complex frequency and the velocity of
the second sound respectively. The latter is determined by the expression
\begin{equation*}
  u_2=\frac{S}{\rho^{1/2}} \sqrt{\frac{\partial T}{\partial S}\frac{\rs}{\rn}} \propto \tau^{1/3}.
\end{equation*}
Further assuming that at the critical region
boundary all terms in \eqref{damping} make contribution
of the same order to the second sound damping we get for the kinetic
coefficients
\begin{equation}
\label{zeta-crit}
 \eta \propto \z_1 \propto \z_2 \propto \z_3
\ \propto \ 
\rc u_2 \rs^{-1}
\ \propto \ 
 \tau^{-1}.
\end{equation}

To find the main term of the space derivatives in \eqref{rho} we note
that 
\begin{equation*}
  \pd{}{z} \propto
  \rho g \left(\frac{\partial}{\partial p}\right)_T \propto
-  \frac{\rho g}{T} \frac{\D T_\lambda}{\D p} \left(\frac{\partial}{\partial \tau}\right)_p.
\end{equation*}
Here the derivative $\D T_\lambda/\D p$ is taken along the phase
transition curve. Substituting this in \eqref{rho} and using
\eqref{zeta-crit} we get\footnote{It should
be emphasized that our accuracy does not allow to resolve between
$T_\lambda$ and $T_\text{c}(Q)$ introduced by some investigators\mycite{exp2}.
In linear approximation these two temperature are identical.}
\begin{equation}
\label{rho-critical}
  \varrho \propto \frac{g^2}{\tau^2}.
\end{equation}

Having shown that finite temperature gradient in superfluid may exist,
it is instructive to find the maximum gradient possible. Obtained
results on effective heat resistance (being derived within linear
approximation) correspond to small temperature gradient and
velocities. They are therefore not applicable to the critical velocities
region. We will try, however, to use them as an estimate beyond the
limits. It is shown in Ref.~\onlinecite{AM} that the critical energy
flux $Q_\text{c}$ in the vicinity of the $\lambda$-transition behaves roughly as $Q_\text{c}
\propto \tau^{4/3}$. Using \eqref{rho-critical}
we see that the maximum temperature gradient $T'_\text{c}$ increases when the
temperature approaches $T_\lambda$ as $T'_\text{c}\propto\tau^{-2/3}$.

\section*{ACKNOWLEDGMENTS}
The author is indebted to A.F.Andreev for attention to this work and
to R.V.Duncan and D.A.Sergatskov for fruitful discussions.  This work
was supported in parts by INTAS grant 01-686, CRDF grant
RP1-2411-MO-02, RFBR grant 03-02-16401, and RF president program.


\begin{thebibliography}{9}
\bibitem{LL6}
{L.D. Landau, E.M. Lifshitz, \textit{Fluid Mechanics}
(Pergamon Press, Oxford, 1987).}
\bibitem{khalat}
{I.M. Khalatnikov, \textit{An Introduction to the
Theory of Superfluidity}. (W.A. Benjamin, New York-Amsterdam
1965).}
\bibitem{exp1}
{{D.A. Sergatskov, A.V. Babkin, S.T.P. Boyd, R.A.M. Lee, R.V. Duncan},
\textit{JLTP} \textbf{134}, Nos.1/2, 517 (2004).}
\bibitem{PP}
{A.Z. Patashinskii, V.L. Pokrovskii \textit{Fluctuation theory of phase
  transitions}. (Oxford; New York: Pergamon Press, 1979).}
\bibitem{alpha}
{J.A. Lipa, J.A. Nissen, D.A. Stricker, D.R. Swanson, T.C.P. Chui,\\
\href{http://www.arxiv.org/abs/cond-mat/0310163}{\texttt{cond-mat/0310163}}.}
\bibitem{joseph}
{B.D. Josephson, {\it Phys. Lett.} \textbf{21}, 608 (1966).}
\bibitem{exp2}
{{R.V. Duncan, G. Ahlers, and V. Steinberg}, \textit{Phys. Rev. Lett.}
\textbf{60} 1522 (1988).}
\bibitem{AM}
{A.F. Andreev, L.A. Melnikovsky,
\href{http://www.arxiv.org/abs/cond-mat/0405111}{\texttt{cond-mat/0405111}}. To appear in
\textit{JLTP} \textbf{135}, Nos.5/6 (2004).}
\end{thebibliography}
\end {document}